\documentclass[
reprint,
aps,
prresearch,
superscriptaddress,
amsmath,
amssymb,
nofootinbib
]{revtex4-2}
\usepackage[T1]{fontenc}
\usepackage[utf8]{inputenc}
\usepackage[polish, english]{babel}

\usepackage[colorlinks = true, linkcolor = blue, urlcolor = blue, citecolor = red, anchorcolor = blue]{hyperref} 
\usepackage[dvipsnames]{xcolor} 
\usepackage{soul} 

\usepackage{amsmath, amssymb, mathtools}
\usepackage{physics} 
\usepackage{dsfont}
\usepackage{tikzit}

\tikzstyle{CP_Orange}=[ultra thick, fill={rgb,255: red,240; green,170; blue,120}, draw=black, shape=chamfered rectangle, chamfered rectangle corners=north west, chamfered rectangle angle=30, tikzit category={CP_Maps}, inner sep=0, minimum height=.5cm, minimum width=.75cm]
\tikzstyle{CP_Blue}=[ultra thick, fill={rgb,255: red,128; green,217; blue,255}, draw=black, shape=chamfered rectangle, chamfered rectangle corners=north west, chamfered rectangle angle=30, tikzit category={CP_Maps}, inner sep=0, minimum height=.5cm, minimum width=.75cm]
\tikzstyle{ketGrey}=[ultra thick, fill={rgb,255: red,217; green,215; blue,238}, draw=black, shape=kite, shape border uses incircle, shape border rotate=247.5, tikzit category=kets, kite upper vertex angle=135, kite lower vertex angle=45]
\tikzstyle{ketGreyBig}=[ultra thick, fill={rgb,255: red,217; green,215; blue,238}, draw=black, shape=kite, shape border uses incircle, shape border rotate=247.5, tikzit category=kets, kite upper vertex angle=135, kite lower vertex angle=45, minimum width=2cm, tikzit fill={rgb,255: red,178; green,182; blue,238}]
\tikzstyle{discard}=[circuit ee IEC, ultra thick, ground, draw=black, tikzit category=symbols, scale=1.5, rotate=90]
\tikzstyle{CP_Green}=[ultra thick, fill={rgb,255: red,116; green,231; blue,116}, draw=black, shape=chamfered rectangle, chamfered rectangle corners=north west, chamfered rectangle angle=30, tikzit category={CP_Maps}, inner sep=0, minimum height=.5cm, minimum width=.75cm]
\tikzstyle{rot90}=[shape=circle, rotate=90]
\tikzstyle{rot270}=[rotate=270]
\tikzstyle{braYellowBig}=[ultra thick, fill={rgb,255: red,255; green,253; blue,187}, draw=black, shape=kite, shape border uses incircle, shape border rotate=292.5, tikzit category=bras, kite upper vertex angle=135, kite lower vertex angle=45, minimum width=2cm]

\tikzstyle{doubled}=[ultra thick, -, tikzit draw={rgb,255: red,135; green,11; blue,159}]
\tikzstyle{dash doubled}=[ultra thick, -, dashed, tikzit draw=magenta, tikzit fill={rgb,255: red,189; green,127; blue,191}, path fading=north, fill={rgb,255: red,239; green,222; blue,240}]
\tikzstyle{Process_Interior}=[-, fill={rgb,255: red,239; green,222; blue,240}, ultra thick]
\tikzstyle{dashed-doubled}=[-, dashed, ultra thick, tikzit draw={rgb,255: red,128; green,0; blue,128}]

\input{q.tikzdefs}

\newcommand*{\mc}[1]{\ensuremath\mathcal{#1}}

\newcommand*{\DensityOp}{\mc D}
\renewcommand*{\H}{\mc H}
\newcommand*{\LL}{\mc L}
\newcommand*{\C}{\mc C}
\newcommand{\W}{\mathcal W}

\newcommand{\V}{\mathcal V}

\newcommand*{\M}{\mc M}
\newcommand*{\N}{\mc N}
\newcommand*{\I}{\mc I}
\newcommand*{\J}{\mc J}

\newcommand*{\perf}[1][\phantom{.}]{
    \mathbin{\raisebox{3pt}{
        \ensuremath{\overset{#1}{\top}}
    }}
}

\newcommand*{\sites}{\ensuremath\textbf{sites}}

\newcommand*{\outcomes}{\ensuremath\textbf{outcomes}}

\newcommand{\id}{\mathds{1}}

\renewcommand{\st}[1]{}

\begin{document}

\title{Causal Order Cannot Be An Observable.}
\author{Declan Maguire}
    \email{D.Maguire@uq.edu.au}
    \affiliation{School of Mathematics and Physics,\\The University of Queensland,\\St Lucia, QLD 4072, Australia}
\author{Fabio Costa}
    \email{Fabio.Costa@su.se}
    \affiliation{Nordita, Stockholm University and KTH Royal Institute of Technology, Hannes Alfv\'ens v\"ag 12 Stockholm, 106 91, Sweden}
\affiliation{School of Mathematics and Physics,\\The University of Queensland,\\St Lucia, QLD 4072, Australia}

\begin{abstract} 
Recent developments in the formalisation of quantum causal structures have made it possible to test and compare hypotheses about causal structure \textit{empirically}, rather than being a-priori assumptions. Such differences in causal structure may be leveraged to distinguish between the processes they belong to, akin to distinguishing between quantum states known to belong to different eigenspaces of an observable. So how far this analogy can be pushed? Can causal order be interpreted as a kind of observable? Can it be measured?

To this end we construct a completely \textit{operational} definition of observables in terms of `discrimination tasks'. This begins with some base set of classes (analogous to a quantum observable's eigenspaces) before we impose three conditions on how these classes relate under discrimination tasks. These conditions recover the properties of a standard observable, letting us apply this definition to non-state entities such as quantum processes, which naturally encode causal structure.

The first of these conditions (1) asserts that a member of any class is perfectly distinguishable from a member of any other class -- for states, this implies conditions (2) and (3), but we show this does not hold automatically for processes (or indeed channels). By analysing classes of quantum processes having strict causal orders, we find causal classes which are sharp but violate condition 2 (but not 3), and classes which are likewise sharp and violate condition 3 (but not 2). This implies that causal order cannot be an observable.
\end{abstract}

\maketitle

\section{Introduction} 
Traditional formulations of quantum theory presume a fixed causal structure, serving as a background constraint that dictates how systems can exchange information. In stark contrast, in general relativity causal structure is only determined upon solving the field equations, similarly to other dynamical variables. This has motivated formulations of quantum theory in which causal structure is not fixed a priori, but can instead be unknown and discovered empirically \cite{Hardy_2007, Qcorr2012}. Such frameworks are further motivated by foundational questions about causality in quantum theory and practical considerations, where analysis of causal structure can facilitate noise modelling in quantum devices or provide advantages in information-theoretic tasks \cite{Chiribella2012, Araujo2014, Guerin2016, OreshkovGiarmatzi_2016, costa2016, Allen2016, QCausal2019, zych2019, QCausalDiscoveryAlgorithm, parker2021background, Costa2025}.

These developments have fostered the view that causal structure can be regarded as an observable physical property, akin to spin or energy. In particular, the \textit{causal order} between events can be determined experimentally---similar to the value of an observable---and it has been proposed that \textit{superpositions of causal orders} are possible---instantiated by processes such as the `quantum switch' \cite{QComputationsWODefCStruc, Araujo_2015}---in close analogy to superpositions of classically distinct values of an observable. Nevertheless, clues have also emerged that such an analogy might have limitations: unlike ordinary superpositions, the quantum switch relies on a control system determining the order between events on a target system, while a direct superposition of oppositely ordered processes is in fact not possible \cite{Costa2020}. Furthermore, purely classical processes can also exhibit `indefinite causal order' \cite{Baumeler2014PRA, Baumeler_2016, Baumeler2022}, indicating a tension with the idea of quantum superpositions of definite orders.
These conflicting perspectives lead to the question of whether at all causal order can function as an observable or if it should be understood as a different type of entity.

Here we formalise and operationalise this question using the language of \textit{single-shot discrimination tasks} \cite{StateDiscrimination,UnitaryDiscrim,CombDiscrimination}. In this framework, an observable can be associated with a set of classes of states/channels/objects each corresponding to distinct, definite values of said observable. Given a single instance of an unknown object, promised to belong to one of these classes, the task is to determine which class it belongs to empirically. We cast causal order within this broad framework using the process matrix formalism \cite{Qcorr2012}, as it is the most generic way of representing specific causal relationships amongst interventions on quantum systems. However due to the generality of our definition of an observable, our approach immediately extends to any generalised probabilistic theory, and so is not bound to the specifics of quantum theory.

Surprisingly, we find that causal order fails to satisfy two core properties associated with physical observables. First, for finite dimensional systems and a sufficient number $n$ of sites (i.e., `slots' in the process matrix), it is not possible to find a single-shot protocol that allows discriminating between all possible $n!$ orders, even when it is possible to perfectly distinguish \textit{every pair} of differently ordered processes. This is in stark contrast with ordinary observables, whose values are encoded in orthogonal subspaces, and where pairwise discrimination protocols can always be combined into a single projective measurement that distinguishes all possible values.

Furthermore, even for just two sites $A$ and $B$, no protocol can determine the order between them without also specifying the connecting channels. In other words, even when some processes with $A$ before $B$ are perfectly distinguishable from similar process with $B$ before $A$, the two \textit{sets} of oppositely ordered processes cannot generally be distinguished in a single shot. This again departs from ordinary observables, where a single binary measurement can always distinguish between two values of an observable, even for degenerate eigenvalues, where each value is associated with many states

Notably, these absence of these properties is \textit{not} caused by the distinctions between classes failing to be `sharp'; we show this by constructing counterexamples to these two core properties, for which each member of each class is perfectly single-shot discriminable from any member of the others. This is significantly unlike the case for classes of quantum states, where the sharpness and observability of a physical trait necessarily coincide.

Our results indicate that causal order cannot be interpreted as an observable property of quantum processes, but it is rather a global property that can only be reconstructed indirectly from correlations.

\section{Background}
This paper is primarily built on the literature on process matrices \cite{Qcorr2012,Araujo_2015}, with discrimination tasks  \cite{StateDiscrimination,UnitaryDiscrim,CombDiscrimination} taking a crucial secondary role. Frameworks closely related to the process matrix formalism include the general boundary formalism \cite{Oeckl2003318}, higher-order quantum transformations \cite{Chiribella2008supermap, Bisio2019}, multi-time states \cite{Aharonov2009, Silva2017}, entangled histories \cite{Cotler2016}, and superdensity operators \cite{Cotler2017}.

\textit{Process matrix theory} takes quantum states, channels, circuits, and so on, and subsumes them all as instances of `quantum processes'. A quantum process can encode arbitrary state and causal information that relates a set of systems, and may be defined either as a tensor or, via the Choi-Jamiołkowski Isomorphism, as a matrix.

\textit{Discrimination tasks} are defined by a kind of guessing game, where an experimenter is given a mystery sample (from a \textit{known distribution}), tests it, then guesses the sample's identity based on the test outcome. The entities being discriminated may be states, channels, circuits, or in the case of this paper, general quantum processes.

This paper will only concern itself with the simplest case of discrimination tasks; `perfect discriminability', i.e. when a pair of processes may be distinguished experimentally without error or ambiguity -- for example, if our processes represent states, this is true if and only if the states are orthogonal.

Because perfect discriminability, unlike orthogonality, is intrinsically a purely operational property, it can generalise beyond states to arbitrary processes readily. This is particularly important because orthogonality of processes, unlike orthogonality of states, does not correspond to perfect discriminability nor have a fixed operational significance.

    \subsection{Process Matrix Formalism}
    Quantum processes are the most general object assigning probabilities to multiple quantum instruments' outcomes consistently. This naturally includes all possible causal relations among the instruments, hence attempts to `measure causal structure' will take the form of an intervention upon processes.

    Throughout this paper we will equivocate between `quantum process matrices' and `quantum process tensors', calling both simply `processes'. These are two equivalent ways of presenting the same theory linked by the Choi-Jamiołkowski Isomorphism, and this paper will primarily use the tensor form.

    Processes are defined with respect to quantum instruments, which are themselves defined with respect to quantum measurements and states. Below, we recount the standard definitions \cite{NielsenAndChuang} of density operators, channels, measurements, and quantum instruments while introducing this paper's notational choices.
    
        \subsubsection{Density Operators, Channels, and Measurements}
        For arbitrary quantum system $S$ of finite dimension, let the space of density operators over that system be denoted
        \begin{align}
            \DensityOp(S)
            :&=  \left\{ \rho \in \LL(\H^S) \Big|
                \Big( \rho \geq 0
                \Big) \wedge
                \Big( \Tr(\rho) = 1 \Big)
            \right\}
        \end{align}
        where $\LL(*)$ is the set of linear operators over a complex Hilbert space $*$.
        
        Single-shot measurements on density operators are given by `POVMs' -- Positive Operator Valued Measures -- a.k.a `Quantum Instruments'. The outcome probabilities of a POVM $\{\sigma_i \in \LL(\H^S)\}_{i \in \outcomes}$ are given by
        \begin{align}
            P_\rho(i)
            = \left\{\Tr\left(\sigma_i\rho\right)\right\}_{i \in \outcomes}
        \end{align}
        where $\sum_i^\outcomes \sigma_i
        = \id^S$, and $\forall i, \sigma_i \geq 0$. POVMs model measurements where the state is either annihilated or disregarded post-measurement, and as such are maps $\DensityOp(S) \to \mathbb{R}^+_0$ directly from states to their measurement outcome probabilities. However if we had a post-intervention state we wished to keep track of, we'd need to instead consider maps from one state-space to another.
        
        PVMs (Projector Valued Measures, a.k.a. von Neumann measurements) are the special case of POVMs where the operators $\sigma_i$, for all $i$, are projectors. Hence, all $\sigma_i$ must be mutually orthogonal, and jointly span the whole space $*$, for the PVM to be a legitimate POVM. A PVM is to a POVM as a pure state is to a mixed state, and are conventionally identified with `sharp' and `unsharp' observables respectively.
        
        This characterisation of a (sharp) observable is equivalent to the more common definition as a self-adjoint operator $A$; consider a set of projectors onto $A$'s eigenspaces which are indexed by their respective eigenvalues -- distinct eigenspaces are always orthogonal, hence the indexed set is a PVM, and the original operator $A$ is recovered by summing each projector scaled by its eigenvalue.
        
        A (deterministic) quantum channel $\C: S_I \to S_O$ is any linear map from states of an `input' system $S_I$ to states of an `output' system $S_O$ which is `CPTP'; i.e. Completely Positive \& Trace-Preserving -- meaning\\
        $\forall S_I',\,\forall \rho \in \DensityOp(S_I \otimes S_I')$,
        \begin{align}
            [\C \otimes \id](\rho)
            &\geq 0
            \intertext{(i.e. CP -- `completely positive'),}
            \Tr\left([\C \otimes \id^{S'}](\rho)\right) &= \Tr(\rho)
        \end{align}
        (i.e. TP -- `trace preserving').
        
        A quantum instrument $\{\M_{i}: S_I \to S_O\}_{i}$ is defined by three properties --
        \begin{equation}
            \sum_{i\in \outcomes}
            \M_{i} \text{ is CPTP,}
        \end{equation}
        \begin{align}
            \forall i &\in \outcomes,&
            &\M_{i} \text{ is CP,}\\
            \forall S_I',\forall \rho &\in \DensityOp(S_I \otimes S_I'),&
            P_\rho (i) = &\left\{\Tr(
            [\M_{i}\otimes \id^{S'}](\rho))\right\}_{i}.
        \end{align}
        
        For succinctness, we will often denote instruments (and other indexed sets) by
        \begin{align}
            \M_{\{i\}} = \{\M_{i}\}_{i \in \outcomes}.
        \end{align}

        While density operators and POVMs are defined with respect to single systems, general quantum instruments are always associated with a \textit{pair} of systems -- in this case, $S_I$ and $S_O$. This paper will henceforth call this pair,
        \begin{equation}
            S_I \otimes S_O = S, \text{ the `site' of the instrument.}
        \end{equation}
        Relative to a site $S$, we say that any instrument $\M_{\{i\}}: \DensityOp(S_I \otimes S_I') \to \DensityOp(S_O \otimes S_O')$ is an `extended instrument' if $S' := S_I' \otimes S_O'$ is non-trivial. When we have multiple instruments $\M^A_{\{a\}}, \M^B_{\{b\}}, \M^C_{\{c\}}\dots$ at sites $A, B, C\dots$, their tensor product $\M^A_{\{a\}} \otimes \M^B_{\{b\}} \otimes \M^C_{\{c\}} \otimes \dots$ is called a `product instrument'\footnote{Note that while a product of instruments is always a valid instrument, a product of \textit{sites} is not a site in general.}.
        
        \subsubsection{Processes}
        If we want to consider \textit{product instruments} and compute their joint probabilities, our answer will depend on how the various constituent instruments relate to each other causally. The standard born rule provides several in-equivalent expressions for the joint probability $P(a,b)$ of a pair of outcomes at sites $A$ and $B$--
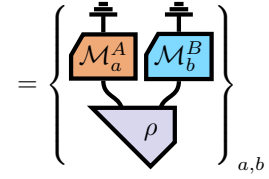
        \begin{align}
            P_{A \parallel B} (a,b) &=
                \Tr\left[
                    \M_{\{a\}}^A \!\otimes\!
                    \M_{\{b\}}^B (\rho)
                \right] \\
            &=
            \left\{
                \begin{tikzpicture}
	\begin{pgfonlayer}{nodelayer}
		\node [style={CP_Orange}] (0) at (-1, 0.75) {$\mathcal M _a^A$};
		\node [style={CP_Blue}] (1) at (1, 0.75) {$\mathcal M _b^B$};
		\node [style=ketGrey] (2) at (0.25, -1.25) {$\rho$};
		\node [style=discard] (3) at (-1, 2) {};
		\node [style=discard] (4) at (1, 2) {};
		\node [style=none] (5) at (-0.5, -0.75) {};
		\node [style=none] (6) at (0.5, -0.75) {};
	\end{pgfonlayer}
	\begin{pgfonlayer}{edgelayer}
		\draw [style=doubled] (1) to (4);
		\draw [style=doubled] (0) to (3);
		\draw [style=doubled, in=-90, out=90, looseness=1.25] (6.center) to (1);
		\draw [style=doubled, in=-90, out=90, looseness=1.25] (5.center) to (0);
	\end{pgfonlayer}
\end{tikzpicture}
            \right\}_{a,b} \nonumber
        \shortintertext{when the instruments are applied to parallel systems, and}
            P_{A<B}(a,b) &=
                \Tr\left[
                    \M_{\{b\}}^B \!\circ\!
                    \M_{\{a\}}^A(\rho)
                \right]\\
            &=
            \left\
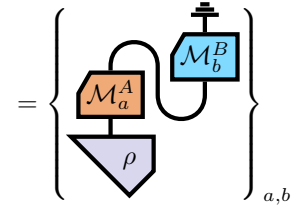
{
                \begin{tikzpicture}
	\begin{pgfonlayer}{nodelayer}
		\node [style={CP_Orange}] (0) at (-1.25, 0.25) {$\mathcal M _a^A$};
		\node [style={CP_Blue}] (1) at (1.25, 1.25) {$\mathcal M _b^B$};
		\node [style=ketGrey] (2) at (-0.75, -1.5) {$\rho$};
		\node [style=discard] (4) at (1.25, 2.5) {};
		\node [style=none] (5) at (-1.25, -1) {};
		\node [style=none] (6) at (1.25, 0.25) {};
		\node [style=none] (7) at (0, 1) {};
		\node [style=none] (8) at (-1.25, 1) {};
		\node [style=none] (9) at (0, 0.25) {};
	\end{pgfonlayer}
	\begin{pgfonlayer}{edgelayer}
		\draw [style=doubled] (1) to (4);
		\draw [style=doubled] (0)
			 to (8.center)
			 to [in=90, out=90, looseness=1.75] (7.center)
			 to (9.center)
			 to [in=-90, out=-90, looseness=1.75] (6.center)
			 to [in=-90, out=90] (1);
		\draw [style=doubled, in=-90, out=90] (5.center) to (0);
	\end{pgfonlayer}
\end{tikzpicture}
            \right\}_{a,b} \nonumber
        \shortintertext{or for the opposite ordering}
            P_{B < A}(a,b) &=
                \Tr\left[
                    \M_{\{a\}}^A \!\circ\!
                    \M_{\{b\}}^B(\rho)
                \right]\\
            &=
            \left\
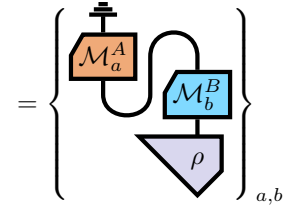
{
            \begin{tikzpicture}
	\begin{pgfonlayer}{nodelayer}
		\node [style={CP_Orange}] (0) at (-1.25, 1.25) {$\mathcal M _a^A$};
		\node [style={CP_Blue}] (1) at (1.25, 0.25) {$\mathcal M _b^B$};
		\node [style=ketGrey] (2) at (1.25, -1.5) {$\rho$};
		\node [style=none] (4) at (1.25, 1.25) {};
		\node [style=none] (5) at (-1.25, 0.25) {};
		\node [style=none] (6) at (1.25, -1) {};
		\node [style=none] (7) at (0, 1.25) {};
		\node [style=discard] (8) at (-1.25, 2.5) {};
		\node [style=none] (9) at (0, 0.25) {};
	\end{pgfonlayer}
	\begin{pgfonlayer}{edgelayer}
		\draw [style=doubled] (1)
			 to (4.center)
			 to [bend right=90, looseness=1.75] (7.center)
			 to (9.center)
			 to [bend left=90, looseness=1.50] (5.center)
			 to [in=-90, out=90] (0);
		\draw [style=doubled, in=-90, out=90] (6.center) to (1);
		\draw [style=doubled] (0) to (8);
	\end{pgfonlayer}
\end{tikzpicture}
            \right\}_{a,b} \nonumber
        \end{align}
        when the instruments are composed sequentially. While this is fine for the majority of purposes in physics, it is inadequate for comparing varied causal structures on a common footing. Quantum processes solve this ambiguity by sidestepping it; a process $W^{ABC\dots}$ is a linear map $\M^A_a \otimes \M^B_b \otimes \M^C_c \dots \to \mathbb{R}$ assigning joint probabilities \textit{directly}. This also makes a process a dual tensor\footnote{Or matrix via the Choi-Jamiołkowski Isomorphism} to products of CP maps.
        
        For such tensors this paper will employ a notation \cite{Hardy2012} inspired by the Einstein summation convention. Suppressing the tensor product symbols $\otimes$ for brevity, a `site index' that is repeated like 
        \begin{equation}    
            \W_{ABC\cdots}[\M^A_a \M^B_b \M^C_c \dots]
        \end{equation}
        shall be contracted, where an upper index denotes maps at a site and a lower index is a `slot' or `hole' that the map is `plugged into'.
    
        Formally \cite{Araujo2017purification}, a process $\W_{ABC\dots}$ is defined as any tensor such that for all sets of extended channels over its sites $\{\C^{\chi \chi'} \in CPTP(\chi\chi')\}_{\chi \in \{A,B,C,\dots\}}$,
        \begin{align}
            \W_{ABC\dots} \left[\C^{AA'}\C^{BB'}\C^{CC'}\dots\right] \in CPTP(A' B' C' \dots),
        \end{align}
        which is necessary and sufficient conditions to ensure positive, normalised probabilities given any choices of instruments and ancillary systems $A_I', A_O', B_I', B_O'\dots$.

        Surprisingly, this is not the most general class of single-shot intervention upon processes guaranteed to have valid probabilities. The most general class are `non-signalling instruments' $\I_{\{i\}}: A_I B_I C_I \dots \to A_O B_O C_O\dots$, whose defining property is that their sum is a `non-signalling channel' -- i.e. $\forall \chi \in \{A,B,C,\dots\}$,
        \begin{align} \label{nosignal}
        \sum_i \Tr_{\chi_O}\I_i (\rho) = \sum_i \Tr_{\chi_O}\I_i \circ \dfrac{\id^{\chi_I}\otimes\Tr_{\chi_I} (\rho)}{d^{\chi_I}}.
        \end{align}
        In practice this means any channel where any site's input cannot signal to another site's output except its own -- and thus, if a site's output is being discarded, its input may also be discarded without effect.

        \subsubsection{Causal Structure}
        We've thus far left causal structure undefined, implicitly given by the way instruments are composed with one-another. Formally, causal influence in processes is defined by inter-site \textit{signalling} -- or more specifically, a lack of causal influence is defined by a lack of signalling.

        Let $A,B,\dots \in \sites$ have a process $\W_{AB\dots}$. We say that `$B$ does not signal $A$' or that `$\W$ is compatible with $A$ preceding $B$' (denoted $A \prec B$), when
        \begin{align}
            \forall \bigotimes_{\chi \in \sites /\{A,B\}} \M^\chi \in \bigotimes_\chi^{} CPTP (\chi)\nonumber\\
            \forall \M^A_{\{a\}}, \forall\M^B, \N^B \in CPTP(B), \nonumber\\
            \forall a, \quad
            \W_{AB\dots}\left[\M^A_{a} \M^B \bigotimes_\chi\M^\chi\right] \\
            =
            \W_{AB\dots}\left[\M^A_{a} \N^B \bigotimes_\chi\M^\chi\right]. \nonumber
        \end{align}
        In words, this means that the outcome probabilities of any instrument at site $A$ are independent of the choice of instrument at site $B$, irrespective of the choices of instruments at all other sites. Note that this is independent of the choice of instrument at $B$, \textit{not} the \textit{outcome} of an instrument at $B$ -- that would be a lack of correlation altogether, not just a lack of signalling/causal influence specifically.
        
        Signalling can be surprisingly subtle; for example, one site might signal to a pair of others in an entangled basis, such that local measurements cannot receive signals but entangled measurements can. It may also happen that inter-site signalling is conditional on the inputs from another site, such as in the quantum switch or its classical analogues. This paper will sidestep these issues by only considering processes that have fixed causal orderings and are markovian -- as we will be using these processes as parts of proofs-by-contradiction, this will not affect the generality of our results.

\section{Setting Up The Problem}
As mentioned in the background section, observables are generally identified with one of two things -- PVMs for `sharp observables', and POVMs for `unsharp observables', with PVMs being equivalent to self-adjoint operators. When we seek to formalise the question of whether causal order is an observable, we mean a `sharp observable' like spin, capable of taking on definite values which may be `plainly measured'.

Generalised beyond states, unsharp observables (POVMs) simply correspond to any valid, outcome-yielding intervention on a quantum entity -- so henceforth, we will simply refer to them as `interventions'. As mentioned with Eq.~\eqref{nosignal}, the broadest category of single-shot interventions upon processes are given by non-signalling instruments.

However, sharp observables (PVMs) -- henceforth simply `observables' --  require a more subtle approach to be generalised. \textit{For this paper we consider the key operational property of an observable is that it maps values to classes of entities, such that a single intervention can identify the value/class of any member entity.} In the case of PVMs, these classes are exactly the eigenspaces of each projector, so to know the classes is to be able to construct the intervention, and vice versa. However if we are to generalise this beyond quantum states to the quantum processes which embody causal structures, we must formalise these properties in purely operational terms.

Concretely, if $\alpha_{jk}$ denotes any member of any such class (members indexed by $k$, classes indexed $j$), then our classes \textit{are observable} when there exists an intervention $\I_{j'}$ such that
\begin{equation} \label{classifier}
    \I_{j'}(\alpha_{jk}) = \delta_{jj'}.
\end{equation}
We can call this the `delta definition' of an observable. Note that it immediately implies that our classes are disjoint.

This delta definition can further be decomposed into three conditions defined in terms of single-shot discrimination tasks:
\begin{enumerate}
    \item If a pair of entities belong to different classes, they are perfectly discriminable,
    \item If all members of some set are sampled from different classes, then a \textit{single} intervention can perfectly discriminate all-from-all jointly,
    \item If an entity is perfectly discriminable from all members of a class, then a \textit{single} intervention can distinguish that entity from the whole class.
\end{enumerate}
Condition 1 here may also be called `sharpness' -- i.e. `sharp classes' are those obeying condition 1 without conditions 2 and 3 necessarily.

When considering only states, sharpness implies conditions 2 and 3 (as condition 1 is there equivalent to orthogonality, and thus our classes are perfectly distinguishable by a PVM), and hence the whole of the delta definition of an observable. But as we will see, this fails when we move beyond states to channels and processes.

By construction, for any observable, any coarse-graining of the classes, or removal of classes, or restriction of class members, will also yield an observable. Conversely, if any combination of coarse-graining, class removal, or member restriction yields a non-observable, the original set of classes will also fail to be an observable.

We will prove that causal order cannot be an observable by constructing classes of processes partitioned by their causal order, for which any sensible definition of `causal classes' would be assigned distinct values, and showing explicitly that they obey condition 1 (they are sharp) while violating conditions 2 and 3. Hence, causal order cannot be an observable. This operational non-equivalence of sharpness and observability in the general case, and its unavoidable presence in causally defined classes, is fundamentally the core result of this paper.

It will be useful to introduce notation for perfect distinguishability of a pair of entities -- we will denote that a pair of entities $\alpha, \beta$ are perfectly distinguishable by an intervention $\I_{\{i\}}$ as
\begin{align}
    &\alpha \perf[\I_{\{i\}}] \beta\\
    \shortintertext{or if $\I_{\{i\}}$ is unspecified,}
    \alpha \perf \beta
    \quad&\Leftrightarrow\quad \exists\, \I_{\{i\}} \quad s.t. \quad \alpha \perf[\I_{\{i\}}] \beta.
\end{align}
Perfect distinguishability of $\alpha$ and $\beta$ is satisfied if and only if
\begin{align}
    \forall i,\quad \I_i(\alpha) > 0 \implies \I_i(\beta) = 0.
\end{align}

\subsection{Being an Observable}
Let us motivate the delta definition of an observable before giving a mathematically formal re-statement of its three conditions. Consider states with spin, for example; if any state has been prepared in a spin-up state, then a spin measurement in the up/down basis will yield `up' with probability 1, and will do so irrespective of the state's other properties. In general, each value of an observable is matched to some class of states, and a single measurement can perfectly identify which class some member-state belongs to. Members of any class may be `plainly observed' to belong to that class with a single, unconditional, measurement -- without error or ambiguity.

Now let us restate our three conditions for an observable: let $\W_{\{j\}\{k\}}$ denote classes (indexed by $j$) whose members (indexed by $k$, independently per $j$) we shall specify as processes without loss of generality. $\W_{\{j\}\{k\}}$ is thus a process observable when:
\begin{enumerate}
    \item \label{Cond.1} $\forall j, j', k, \quad \left(j \neq j' \implies \W_{jk} \perf \W_{j'k}\right)$, \quad\text{(i.e. `sharpness')}
    \item \label{Cond.2} $\forall k, \Big(\forall j, j' \neq j, \exists\, \I_{\{i\}} \quad \Big| \quad \W_{jk} \perf[\I_{\{i\}}] \W_{j'k}\Big)$\\
    $\implies \Big(\exists\, \J_{\{i\}}, \forall j, j' \neq j \quad \Big| \quad \W_{jk} \perf[\J_{\{i\}}] \W_{j'k}\Big)$,
    \item \label{Cond.3} $\phantom{\implies}\Big(\forall k, \exists \I_{\{i\}} \quad\Big|\quad \V \perf[\I_{\{i\}}] \W_{jk}\Big)$\\
    $\implies \Big(\exists \J_{\{i\}}, \forall k \quad\Big|\quad \V \perf[\J_{\{i\}}] \W_{jk}\Big) =: {\Big(\V \perf \W_{j\{k\}}\Big)}$.
\end{enumerate}
We see here that conditions 2 and 3 take on a particularly simple form, essentially that our quantifiers commute. As the reverse implications always hold, we could just as well use $\Leftrightarrow$ to emphasise the symmetry. It is trivial to see that these three conditions are each implied by the delta definition of an observable, Eq.~\eqref{classifier}; the converse, however, is a little more involved to prove.

Firstly note that condition 3 is equivalent to perfect discriminability being preserved under convex sums of class elements (by the linearity of interventions generally), and hence augmenting a class with mixtures makes no difference to which conditions the class satisfies. We may exploit this by constructing representatives of each class $j$ as convex sum of class elements, by having each representative have support over all elements of its (augmented) class
\begin{equation}
    \Bar{\W}_{j*} :\in \left\{\sum_k p_k \W_{jk} \quad\Big|\quad \forall k,\: p_k>0 \wedge \sum_k p_k = 1 \right\}.
\end{equation}
We can call these representatives, because an entity is perfectly discriminable from all our class members if and only if it is perfectly discriminable from such a representative. By sharpness, we hence know all representa are perfectly discriminable pairwise.

Then by condition 2, we can stitch these pairwise discriminations into a global intervention which perfectly distinguishes \textit{all} representa, hence all members of any class from those of any other, thus satisfying Eq.~\eqref{classifier}. This means we can be confident that our three conditions are indeed logically equivalent to Eq.~\eqref{classifier}, itself equivalent to the standard definition of a (sharp) observable over quantum states. As we will later show, conditions 2 and 3 are also independent of one another and are not implied by condition 1, ensuring our three conditions are all logically necessary and non-redundant.

\subsubsection{Closure Under Mappings}
One may wonder whether these conditions may be bypassed by mapping our entities from one space to another, such that classes which violate them might be mapped into another entity space where their images do not violate them. However, this is not so. This can be recognised by noting that such a map\footnote{Assuming such a map to be linear convex, ensuring compatibility with the probabilistic interpretation of convex sums of entities.} from entities of space $X$ to entities of space $Y$ may be reinterpreted as a map on the \textit{dual} space, taking interventions on space $Y$ to interventions on space $X$. This is akin to the way that a linear operator may be seen as either having an action on kets, or equivalently as having a dual action on bras. In this way, any attempt to remap processes to something else before applying some intervention is equivalent to embedding interventions on the transformed space into interventions on the original, \textit{untransformed} space.

Hence any solution for satisfying conditions 2 and 3 in the image space canonically solves the problem on the original space by taking the dual action of our map on interventions. Thus the image of classes of entities may obey the same conditions as the original classes, or a subset of them, but never more. Hence, \textit{failure} to be an observable is closed under convex-sum respecting maps.

This point, while subtle, undergirds the generality of this paper's result and the confidence with which it may be said universally. The failure of causal order to be an observable is not merely due to our choice of mathematical framework, but is intrinsic to the task. In short, it guarantees that no physical map exists from non-observable classes to mutually orthogonal quantum state spaces, no matter how clever the implementation.

\subsection{Encoding Outcomes}
We have thus far been considering interventions on processes yielding a classical outcome. However, for any intervention $\I_{\{i\}}$, we can encode its outcomes in an orthonormal basis $\{\ketbra{i}{i}\}_i$ of some Hilbert space in the joint future of all the process's sites, as 
\begin{equation}
    \dot \I :=\sum_i \ketbra{i}{i} \otimes \I_i.
\end{equation}
This mapping is reversible by a corresponding projective measurement on this new Hilbert space, and crucially, this encoded intervention from processes to a Hilbert space is linear.

Henceforth we will call this extra Hilbert space the `result space'. While this change in representation of an intervention may seem arbitrary, it is pivotal for proving that causal classes can violate condition 2 of being an observable.

Other encodings of outcomes are possible, in particular, coherent or semi-coherent analogues may be defined, from which our chosen incoherent embedding may be recovered by applying a de-cohering channel to our chosen result space. However they provide no benefit for the purposes of this paper, and so we choose the straightforward incoherent mapping for simplicity; our following arguments are independent of the precise features of the channel constructed, so long as some POVM on the outcome space exists that recovers the intervention $\I_{\{i\}}$ when applied. 

\subsection{Causal Classes}
To apply this definition of an observable to causality, we must first have a set of `causal classes'. While attempting to classify all exotic causal structures is beyond the scope of this paper, processes that have a simple causal \textit{order} are much more tractable.

We will further restrict our consideration to processes which are \textit{markovian}, implying that causal influence on a site is mediated entirely by the site immediately preceding, and hence our processes consist of channels from each site to the next, with an initial state for fed into the first site and the output of the final site being discarded \cite{costa2016, Pollock2018, Pollock2018a}. These processes have very simple causal structures, making them a sort of minimal test for whether a causal order observable can exist; no purported causal order observable could rightly be called if it could not successfully distinguish these minimal cases.

So despite these restrictions to special processes, they will not impact the generality of this paper's results! As mentioned below the definition of an observable, any coarse-graining of its classes, removal of their elements (or of classes altogether), must \textit{always result in another observable}. Thus, by disproving that these simple causal classes fail to be observables, we know that any more complicated definition, if generalising these classes, will \textit{also} not be an observable.

However this alone is insufficient, as these markovian ordered processes may become arbitrarily noisy -- and hence arbitrarily similar to a totally mixed process -- without change in causal order until the noise weight is 1. These would fail to be (sharp) observables from the start simply by failing to be sharp at all, in a very boring way\footnote{and, modelling noisy observables is beyond the scope of this paper.}!

To avoid modelling noisy observables, we will require that each site in a process have at least one noiseless bit of capacity to the subsequent site. We may consider these to be `causally strict' -- due to allowing deterministic control of an instrument's outcome via our choice of instrument at the site prior.

Specifically (where $B$ is the immediate successor of $A$), there must exist instruments $\M^A_{\{|x\}}$ and $\N^B_{\{b\}}$ such that 
\begin{equation}
    P(b|x) = \W_{AB}\M^A_{\{|x\}}\N^B_{\{b\}} =  \delta_{\{b , x\}},
\end{equation}
where $b$ and $x$ can take at least two values, and the bar in $|x$ indicates that we are conditioning $b$ (an experimental outcome) on $x$ (an instrument setting that we freely choose). In the markovian, ordered process case, causal strictness reduces to all inter-site channels having a $\geq 1$-bit capacity. In particular, all Markovian processes that consist of an initial state and a unitary connecting each site to the next have a strict causal order.
    
\section{Results}
In this paper we will construct two restricted sets of sharp causal classes, which may be understood as two special cases of sharp causal classes in general, that each obey condition 1 of an observable while violating conditions 2 and 3 respectively. The first counterexample will have a single process per class, with one class per causal order for $n$ sites, while the second will have multiple class members but only two sites, and hence two classes, one per ordering.

    \subsection{Permuted Processes}
    Let us give a mathematically concrete definition of the types of classes of process we will be working with, and prove that they obey condition 1 of an observable.
    
    Consider processes with $n$ isomorphic sites $\{A, B, C\dots\}$ (i.e. all site input spaces are of the same dimension, and likewise all outputs) listed alphabetically, permutation functions $\pi \in \Pi_n$ that may act on (isomorphic) sites in a list to shuffle them, and some `starting' process $\W_*$ which is causally sharp and markovian. These naturally generate an ensemble 
    \begin{equation}
        \W^{ABC\dots}_{\{\pi\}} := \{\W_*^{\pi(A)\pi(B)\pi(C)\dots}\}_{\pi \in \Pi_n}.
    \end{equation}

    Let us define a particularly simple `starting' process
    \begin{align}\label{Wstar}
        \W_* =& 
        [\id^{A_I}/d] \\
        &\otimes \C_{\id}^{A_O \to B_I} \otimes \C_{\id}^{B_O \to C_I} \otimes \cdots \otimes \C_{\id}^{Y_O \to Z_I} \nonumber \\
        &\otimes\Tr_{Z_O} \nonumber
    \end{align}
    where $\dim(A_I) = \dim(A_O) = d$, all sites are isomorphic, $\id^{A_I}/d$ is a totally mixed state at the first site's input, $\C^{\alpha \to \beta}_{\id}$ denotes an identity channel from system $\alpha$ to system $\beta$, and $\Tr_{Z_O}$ discards the output of the final site $Z$.

    Like this starting process, all its permutations are markovian and causally sharp, and all intermediate channels $\C^{\alpha \to \beta}_{\id}$ have the capacity to transmit a bit losslessly. This also holds for any choice of unitary intermediate channels in the starting process.

    Any two such processes (let them be $\W, \W'$) generated by distinct permutations ($\pi, \pi'$) are easily discriminated. Choose any pair of sites ($A,B$) ordered oppositely between the two processes (e.g. $\W$ has $A \prec B$, and $\W'$ has $B \prec A$), and fill every site other than $A$ and $B$ with identity channels\footnote{If the intermediate channels are non-unitary, then a more complex choice of instruments may be needed that maps the incoming bit to the outgoing bit}. Each instrument at $A$ and $B$ then attempt to measure an incoming bit in the appropriate basis for our intermediate channels, before sending the same bit onwards either as-is (say, at site $A$) or flipped (at $B$), again in the appropriate basis. If the measurements from each site in the pair match, we know the non-flipping site (i.e. $A$) came first, conversely, if they differ then the flipping site (site $B$) was first.

    \textbf{Therefore, permuted processes defined in this way will satisfy condition 1.}

    This protocol, specifically once the problem is reduced to two remaining sites, is found verbatim in \cite{lewandowskastrategies2022} for the unitary channel case.

    \subsection{Violating Condition 2}
    We know already that a single-shot intervention $\I_{\{i\}}$ on processes may be recast as a linear map $\dot\I$ from processes to a `result space'.

    The image of any linear map is always of equal or lower dimension than its domain. For $\dot\I$, its domain is the processes over the sites of $\W$, which is itself a subset of all \textit{tensors} over those sites. We use this to derive a loose bound on the dimension of $\dot\I$'s image; the Hilbert space dimension of each site input and output is identically $d$ by construction, and so each site contributes a factor of $d^2 \cdot d^2 = d^4$ to the dimension of the overall tensor space, meaning \textbf{the dimension of $\dot\I$'s image is less than or equal to $d^{4n}$}.

    When we consider the image $\dot\I (\W_{\{\pi\}}) = \rho_{\{\pi\}}$, we know that it has $n!$ elements, one per permutation. However for any value of $d$ there exist $n$ for which
    \begin{align}
        n! > d^{4n}.
    \end{align}
    The size of an orthonormal set cannot exceed the ambient tensor space in which it is found, and hence any oversized set must have non-orthogonal members. 

    \textbf{Therefore, no intervention $\dot\I$ exists taking all permuted processes $\W_{\{\pi\}}$ to orthogonal states, and so causal classes violate condition 2 of being an observable}.

    Looking at the result from the other direction, the minimal dimension to discriminate all causal orders for $n$ sites must satisfy 
    \begin{equation}\label{localdimbound}
        d\geq  (n!)^{\frac{1}{4n}} \approx n^{\frac{1}{4}}.
    \end{equation}
    For $d = 2$, this bound requires $n \approx 41$ in order to be broken, with small increases in $d$ leading to rapid increases in the minimum $n$ which violates the bound. For any finite $d$, this minimum $n$ is finite and fixed -- and conversely, for $n$ sites to (possibly) admit a causal order observable, $d$ cannot be too small. However, we know this bound is not tight (as a space of valid processes is strictly smaller than its ambient tensor space, whence came this bound), although whether perfect discrimination protocols always exist when the space of processes is smaller than $n!$ remains unsolved.

    We may note that because each restricted class here contains only a single element and satisfy condition 1 of an observable, they trivially satisfy condition 3, showing that given condition 1, condition 3 does not imply condition 2.

    \subsubsection{Regarding Channels}
    It is not in fact mandatory that our entities be quantum processes for them to violate condition 2 whilst obeying conditions 1 and 3; we can achieve the same outcome with channels. Let us define singlet classes of channels
    \begin{equation}
        \C_{\{\pi\}}(\rho) := \left\{\sum_i^d \ketbra{\pi(i)}{\pi(i)} \otimes \bra{i}\rho\ket{i}\right\}_{\pi}.
    \end{equation}
    Any two distinct channels shall correspond to distinct permutations of the computational basis, and so are easily discriminated by feeding in a state they permute differently and then measuring the outcome in the computational basis.

    Single-shot discrimination tasks on quantum channels are optimised when given access to an ancilla of dimension no less than the channel input; greater dimensions provide no advantage. This implies that the largest effective Hilbert space available for measurement is of dimension no greater than the square of the channel input dimension. By an essentially identical argument to the above we can thus conclude no input, with an ancilla or otherwise, will be mapped to an orthonormal set by the $d!$ permuting channels.

    \subsection{Violating Condition 3}
    Let us now consider the case $n = 2$. Here $\W_*$, Eq.~\eqref{Wstar}, has only two permutations, so we will adopt $\W$ for the $A \to B$ unpermuted case and $\V$ for the $B \to A$ permuted case. Clearly $\W \perf \V$ for the same reasons given in the prior section.

    For the sake of making explicit computation easy, let us also fix $d=2$ as the dimension of site inputs and outputs\footnote{In the general $d$ case we simply use the Pauli-Weyl operators, a $d^2$ sized orthogonal set of unitary operators, to generate a qudit depolarising channel; see \ref{PauliWeyl} for proof}. As we are only seeking a counterexample, this will not harm our overall argument. For each causal order, we consider four processes connecting the two sites through each of the Pauli operators:
    \begin{align}
        \W_{\{j \}} &= \{\W_\id, \W_X, \W_Y, \W_Z\},\\
        \V_{\{j'\}} &= \{\V_\id, \V_X, \V_Y, \V_Z\},
    \end{align}
    \begin{subequations}
    where
    \begin{alignat}{8}
        \W_\id &=& [\id^{A_I}/d] \otimes \C_{\id}^{A_O \to B_I} \otimes\Tr_{B_O},\\
        \W_X   &=& [\id^{A_I}/d] \otimes \C_{X  }^{A_O \to B_I} \otimes\Tr_{B_O},\\
        \W_Y   &=& [\id^{A_I}/d] \otimes \C_{Y  }^{A_O \to B_I} \otimes\Tr_{B_O},\\
        \W_Z   &=& [\id^{A_I}/d] \otimes \C_{Z  }^{A_O \to B_I} \otimes\Tr_{B_O}\\
        \shortintertext{and}
        \V_\id &=& [\id^{B_I}/d] \otimes \C_{\id}^{B_O \to A_I} \otimes\Tr_{A_O},\\
        \V_X   &=& [\id^{B_I}/d] \otimes \C_{X  }^{B_O \to A_I} \otimes\Tr_{A_O},\\
        \V_Y   &=& [\id^{B_I}/d] \otimes \C_{Y  }^{B_O \to A_I} \otimes\Tr_{A_O},\\
        \V_Z   &=& [\id^{B_I}/d] \otimes \C_{Z  }^{B_O \to A_I} \otimes\Tr_{A_O}.
    \end{alignat}
    \end{subequations}

    The two sets $\W_{\{j \}}$ and $\V_{\{j'\}}$ each constitute a sharp causal class (one per ordering of two total orderings), and each member of one class is perfectly distinguishable from each member of the other, satisfying condition 1 of an observable.

    We know already that condition 3 of an observable is satisfied only if all convex sums of one class are perfectly distinguishable from all convex sums of the other -- i.e.
    \begin{align}
        \W_{\{j\}} \perf \V_{\{j'\}} \implies \left[\sum_j P(j) \W_j \right] \perf \left[\sum_{j'} P(j')\W_{j'} \right].
    \end{align}
    Let us compute these for equal mixtures:
    \begin{alignat}{8}
        \Bar{\W} :&= \frac 1 4 [&\W_\id& + &\W_X& + &\W_Y& + &\W_Z&],\\
        \shortintertext{and}
        \Bar{\V} :&= \frac 1 4 [&\V_\id& + &\V_X& + &\V_Y& + &\V_Z&].
    \end{alignat}
    Factoring out common terms, expanding these convex sums yields
    \begin{alignat}{8}
        \Bar\W =&
        [\id^{A_I}/d] \\
        &\otimes \frac 1 4 \left[
        \C_{\id}^{A_O \to B_I} +
        \C_{X  }^{A_O \to B_I} +
        \C_{Y  }^{A_O \to B_I} +
        \C_{Z  }^{A_O \to B_I} \right] \nonumber\\
        &\otimes \Tr_{B_O} \nonumber
        \shortintertext{and}
        \Bar\V =&
        [\id^{B_I}/d]\\
        &\otimes \frac 1 4 \left[
        \C_{\id}^{B_O \to A_I} +
        \C_{X  }^{B_O \to A_I} +
        \C_{Y  }^{B_O \to A_I} +
        \C_{Z  }^{B_O \to A_I} \right] \nonumber\\
        &\otimes \Tr_{A_O}. \nonumber
    \end{alignat}
    However, this sum is just the Kraus form of a totally depolarising channel (where the operators are themselves expressed as channels), i.e. for channels $\phi \to \psi$,
    \begin{align}
        \frac 1 4 \left[
        \C_{\id}^{\phi \to \psi} +
        \C_{X  }^{\phi \to \psi} +
        \C_{Y  }^{\phi \to \psi} +
        \C_{Z  }^{\phi \to \psi} \right]\\
        =
        \Tr_\phi \otimes [\id^\psi/d]\nonumber
    \end{align}
    so we can substitute this into our expressions to yield
    \begin{align}
        \Bar\W
        &= [\id^{A_I}/d] \otimes \Tr_{A_O} \otimes [\id^{B_I}/d] \otimes \Tr_{B_O},
        \shortintertext{and}
        \Bar\V
        &= [\id^{B_I}/d] \otimes \Tr_{B_O} \otimes [\id^{A_I}/d] \otimes \Tr_{A_O}\\
        &= [\id^{A_I}/d] \otimes \Tr_{A_O} \otimes [\id^{B_I}/d] \otimes \Tr_{B_O}\\
        &= \Bar\W.
    \end{align}
    But this contradicts our assumption that $\Bar\W \perf \Bar\V$, as no process can be perfectly distinct from itself. \textbf{This implies that $\W_{\{j\}} \perf \V_{\{j'\}}$ must be false, and hence causal classes do not obey condition 3 of an observable.}

    We may note also that this pair of restricted classes obey condition 2 of an observable trivially (as we have only two classes, and hence interventions to discriminate members of a set sampled from \textit{all} classes are exactly the same as those that can discriminate between samples from just \textit{two} classes, which is just sharpness), demonstrating that given condition 1, condition 2 does not imply condition 3.

    \subsubsection{Violating Conditions 2 \& 3 Jointly}
    Thus far we have shown that, given condition 1, conditions 2 \& 3 may be both true (in which case our sharp classes form an observable), or that 2 may be violated while 3 holds, or vice versa (as shown in the above two examples). To show that conditions 2 and 3 are indeed independent given condition 1, it remains to show that sharp classes may violate \textit{both} conditions simultaneously. Thankfully, this can be done by essentially combining our two counterexamples.
    
    Consider the `starting process' $\W_*$ we used to generate our permuted processes -- in it, we decided that the evolution from each site output to the following site input would just be an identity channel for simplicity. We could instead decide to replace this singular starting process with a class of them, constructed by replacing the intra-process identity channels with any choice of Pauli channels (as we did when trying to violate condition 3), resulting in a `starting class' of $4^{n-1}$ processes. We may then apply our site permutations (as we did when trying to violate condition 2) to generate our full set of classes.
    
    Such classes will remain sharp by the previously discussed procedure for discriminating oppositely ordered processes (requiring only that the choice of signalling basis are adapted to the intermediate unitary channels). As these classes are just augmentations of those from subsection B (which were constructed to violate condition 2), and augmenting classes cannot improve their distinguishability, our larger classes must also violate condition 2.
    
    By an essentially identical argument to that in subsection C, we may also cause these classes to collapse into one another by averaging over our intermediate Pauli channels to generate the same effective process out of two distinct classes. \textbf{Therefore, these classes violate condition 3 in addition to 2, while remaining sharp, and so conditions 2 and 3 are independent in general given condition 1.}

\section{But \textit{can} causal order be an observable?}
We have based our analysis on the process matrix formalism, which provides an operational framework for processes without a pre-assigned causal structure, and one may wonder if the notion of causal order as an observable could be recovered in some different framework. It is worth in particular to compare our findings with perspectives arising from general relativity, where the idea of causal order as an observable may arise naturally. Most relevant in this respect is the proposal from Ref.~\cite{delahamette2022quantum} of an observable for causal order between operationally defined events. 

Ref.~\cite{delahamette2022quantum} takes as a starting point the definition of event location as a coincidence of worldlines, an idea introduced by Einstein to address the `hole argument' and provide a meaningful notion of events in a diffeomorphism invariant theory \cite{Stachel2014}. This definition allows for a quantum extension, where worldlines are replaced with single-particle wavepackets, but classical worldines will be sufficient for comparison to our analysis.

The setup in Ref.~\cite{delahamette2022quantum} is the following: a `target particle' moves along a certain worldline and meets two other particles along its journey. The two crossings---where interactions between the particles may occur---define the two events $A$ and $B$ whose causal order is to be determined. This is done by looking at the sign of the proper time difference between the two events along the target's worldline, $\Delta\tau := \tau_B-\tau_A$. If $\Delta\tau>0$, then $B$ is after $A$, while $A$ is after $B$ for the opposite sign. Although formulated in terms of geometrical quantities, this protocol can be made operational by letting a clock travel along the target's worldline and measuring it at the two crossing events. This procedure maps to our process-matrix formulation by identifying the crossing events with the corresponding sites in a process, and the clock with (a subsystem of) the local system at each site. In light of this, the statements in Ref.~\cite{delahamette2022quantum} are fully compatible with our findings: for just two events, where the proper time difference is fixed up to a sign, and where the evolution of the clock along the worldline is a known unitary, observing causal order reduces to distinguishing two processes connected by unitaries in opposite orders, which we have seen is always possible for any system size. However, as soon as any of the above conditions is relaxed, we run into the obstructions from our results.

In detail: if we want to measure the causal order for a larger number $n$ of sites, we have seen that a minimum local dimension is needed, implying a violation of Condition 2.\footnote{To closely reproduce the protocol in Ref.~\cite{delahamette2022quantum}, the clock should be initialised in some fiducial pure state at the first site (rather than in the maximally mixed state as in our example) and the unitary between sites should cycle through orthogonal states. In order to read off all proper times perfectly, the clock's dimension would have to be at least $n$, which is more than the lower bound we found in Eq.~\eqref{localdimbound}.}
Additionally, if the proper time between the events is not fixed in advance, we run into the limitations expressed by the failure of Condition 3. Indeed, fixing the Hamiltonian $H$ driving the clock along the target's worldine, the total unitary $e^{-i H\Delta \tau}$ depends on $\Delta \tau$. Determining the causal order for all possible values of $\Delta \tau$ (within some range), would be equivalent to discriminating across oppositely ordered processes within the whole range of unitaries, which we know cannot be done in general. A similar obstruction arises if the clock's Hamiltonian is not given.

So far, our analysis concerns putative observables of process matrices, namely of a set of quantum degrees of freedom and their correlations. One may object that causal order should not be viewed as a property of such systems, but of a background spacetime. In this view, the quantum degrees of freedom in the process matrix merely represent the \textit{measurement devices} used to observe causal structure---just as clocks are tools to measure time. However, causal order in this context \textit{is not} a physical observable analogous to energy or momentum: it is a property of the background spacetime metric, which is only indirectly accessible. Assessing causal order should then be seen as a parameter estimation task \cite{Helstrom1969} --- albeit with the peculiar property that the parameter is discrete-valued ---and not as the measurement of an observable. Hence, we conclude again that causal order is not an observable in this perspective. But, spacetime \textit{is not} a fixed background in general relativity! It has its own dynamical degrees of freedom, presumably living in a Hilbert space within a candidate theory of quantum gravity, and one may expect causal order to be an observable for those degrees of freedom. However, this view brings us back to our analysis---where indeed we attempt to formalise causal order as an observable on quantum degrees of freedom---and to our main conclusion: that causal order \textit{cannot} be a quantum observable, within a reasonably broad interpretation of the term.

The takeaway of this is that if spacetime is, indeed, a quantized degree of freedom in a future theory of quantum gravity, then ``causal order" cannot be directly measured as a sharp observable akin to energy, momentum, or spin, but must instead be indirectly reconstructed through global quantum correlations.

\section{Conclusions}

Our results counter a recurring narrative that causal order---or more generally causal structure---can be treated \textit{on par} with other physical observables in quantum theory. This does not contradict the idea that causal order can be indefinite, nor that observables can be used to witness causal properties \cite{Araujo_2015, WitnessIntro, Wechs2019, MarkovProcesses}. Rather, our work motivates the development of new, specific methodologies for the study causal structure, with discrimination tasks, and more generally hypothesis testing, emerging as promising directions \cite{QHypothesisTesting, Chiribella2019a, Kundu2022}.

It remains notable the particular way in which causal order fails at being an observable: in specific situations, causal order \textit{does} behave operationally like an observable---specifically, it is possible to perfectly distinguish the causal order between two processes whose connecting unitaries are known. However, such discrimination procedures do not extend to a joint discrimination procedure within a set, even when each pairs within the set are perfectly distinguishable. It is also not possible to extend discrimination between all pairs across two oppositely-ordered sets to a single procedure to discriminate the sets as a whole. Although surprising in comparison to states, we have seen that similar situations can be expected for discrimination of channels, where, just as for processes, perfect distinguishability does not imply orthogonality. This suggests that causal relations cannot be detached from the specific transformations between causal relata: even if a set of processes is associated with a single causal structure, reconstructing such structure---or discriminating it from another one---must be tied to the specific properties of the observed processes. Apart from possible methodological consequences in applied contexts, this indicates that a geometrical notion of causal structure, compatible with general relativity, cannot be maintained in a more fundamental theory where causal relations are encoded in quantum degrees of freedom.

\appendix 
\section{Qudit Depolarising Channel Kraus Operators}\label{PauliWeyl}
The Pauli-Weyl matrices are the qudit generalisation of the Pauli matrices for qubits, a $d^2$ set of unitary matrices $\{X_pZ_q\}_{pq}$ which are orthogonal and of norm $d$. By definition $X_p := \sum_{m=0}^{d^2 -1}\ketbra{m+p}{m}$ (the `shift' matrices, generalising Pauli $X$), and $Z_q := \sum_{n = 0}^{d^2 - 1} \ketbra{n}{n} e^{\frac{2\pi iqn}{d}}$ (the `clock' matrices, generalising Pauli Z). They hold a special commutation relation that $Z_qX_p = e^{\frac{2\pi ipq}{d}}X_pZ_q$.

The claim that the Pauli-Weyl matrices form the Krauss operators of the qudit totally depolarising channel (just as the regular Pauli matrices are in the qudit case) may be expressed by saying that if 
\begin{align}
    \C(\rho)
        :&= \frac{1}{d^2} \sum_{p,q = 0}^{d-1} [X_pZ_q] \rho [X_pZ_q]^\dagger\\
        &= \frac{1}{d^2} \sum_{p,q = 0}^{d-1} X_pZ_q \rho Z_q^\dagger X_p^\dagger
        \intertext{then it follows that}
        \C(\rho) = \Tr(\rho) \cdot \id / d
\end{align}
-- i.e. the qudit depolarising channel. Note that as a mixture of unitary channels, $\C$ must be a valid channel also.

Firstly, recognise that $\rho$ itself, being a $d \times d$ matrix, admits a generic decomposition in the Pauli-Weyl basis as $\rho := \sum_{p',q' = 0}^{d-1} \alpha_{p'q'} X_{p'}Z_{q'}$ for some coefficients $\alpha$. By their definition $X_pZ_q$ are traceless unless $p=q=0$, in which case $X_0Z_0 = \id$ and hence $\Tr(X_0Z_0) = d$, so it follows from the normalisation condition $\Tr(\rho) = 1$ that $\alpha_{00} = 1/d$. Secondly, note that $X_p^\dagger = X_{-p}$ and likewise $Z_q^\dagger = Z_{-q}$, hence $\C(\rho) = \frac{1}{d^2}\sum_{p,q = 0}^{d-1} X_pZ_q \rho Z_{-q} X_{-p}$.

We may now begin the core of the proof. Let us consider the action of $\C$ on an arbitrary (normalised) $\rho$ -- thus,
\begin{align}
    \C(\rho)
    &= \frac{1}{d^2}\sum_{p,q = 0}^{d-1} X_pZ_q \rho Z_{-q} X_{-p}\\
    &= \frac{1}{d^2}\sum_{p,q = 0}^{d-1} X_pZ_q \left[\sum_{p',q' = 0}^{d-1} \alpha_{p'q'} X_{p'}Z_{q'}\right] Z_{-q} X_{-p}\\
    &= \frac{1}{d^2}\sum_{p,q = 0}^{d-1}\sum_{p',q' = 0}^{d-1} \alpha_{p'q'} X_pZ_q X_{p'}Z_{q'} Z_{-q} X_{-p}
    \intertext{By passing the $Z_{-q}$ term through $Z_{q'}$ (which it commutes with) and $X_{p'}$ (picking up a phase), we get}
    &= \frac{1}{d^2}\sum_{p,q = 0}^{d-1} \sum_{p',q' = 0}^{d-1} \alpha_{p'q'} X_pZ_qZ_{-q} X_{p'}Z_{q'} X_{-p} \cdot e^{\frac{-2\pi ip'q}{d}}\\
    &= \frac{1}{d^2}\sum_{p,q = 0}^{d-1} \sum_{p',q' = 0}^{d-1} \alpha_{p'q'} X_p X_{p'}Z_{q'} X_{-p} \cdot e^{\frac{-2\pi ip'q}{d}}
    \intertext{As $Z_{-q} = (Z_q)^{-1}$. Likewise passing $X_{-p}$ through and cancelling we get}
    &= \frac{1}{d^2}\sum_{p,q = 0}^{d-1} \sum_{p',q' = 0}^{d-1} \alpha_{p'q'} X_{p'}Z_{q'} \cdot e^{\frac{-2\pi ip'q}{d}} \cdot e^{\frac{-2\pi ipq'}{d}}\\
    &= \frac{1}{d^2} \sum_{p',q' = 0}^{d-1} \alpha_{p'q'} X_{p'}Z_{q'} \cdot \sum_{q = 0}^{d-1} e^{\frac{-2\pi ip'q}{d}} \cdot \sum_{p = 0}^{d-1}e^{\frac{-2\pi ipq'}{d}}\\
    &= \frac{1}{d^2} \sum_{p',q' = 0}^{d-1} \alpha_{p'q'} X_{p'}Z_{q'} \cdot (d \cdot \delta_{p'}) \cdot (d \cdot \delta_{q'})\\
    &= \sum_{p',q' = 0}^{d-1} \alpha_{p'q'} X_{p'}Z_{q'} \cdot \delta_{p'} \cdot \delta_{q'}\\
    &= \alpha_{00} X_0 Z_0\\
    &= \frac{1}{d} \id \quad \Box.
\end{align}

Note that such a decomposition into unitary Krauss terms is not unique. Because a depolarising channel is invariant under pre- and post-composition with unitary channels, any pre- and post-composition of a unitary with the set of Krauss terms generates another valid set of Krauss terms, meaning any unitary channel may be such a Krauss term in some decomposition by considering the composition of that very unitary with the `canonical' decomposition, and specifically its action on the identity term.

\begin{acknowledgments}
We thank A.-C.~de la Hamette and S.~Shrapnel for useful discussions. We acknowledge support from the Knut and Alice Wallenberg Foundation through the Wallenberg Initiative for Network and Quantum Information (WINQ). This work is supported from COST Action CA23115: Relativistic Quantum Information, funded by COST (European Cooperation in Science and Technology). This research was supported by the Commonwealth through an Australian Government Research Training Program Scholarship [DOI: https://doi.org/10.82133/C42F-K220]. We acknowledge the traditional owners of the land on which the
University of Queensland is situated, the Turrbal and Jagera people.
\end{acknowledgments}

\bibliographystyle{ieeetr}
\bibliography{ProcessBibliography}

\begin{thebibliography}{10}

\bibitem{Hardy_2007}
L.~Hardy, ``Towards quantum gravity: a framework for probabilistic theories with non-fixed causal structure,'' {\em Journal of Physics A: Mathematical and Theoretical}, vol.~40, pp.~3081--3099, mar 2007.

\bibitem{Qcorr2012}
O.~Oreshkov, F.~Costa, and {\v{C}}.~Brukner, ``Quantum correlations with no causal order,'' {\em Nature Communications}, vol.~3, p.~1092, 10 2012.

\bibitem{Chiribella2012}
G.~{Chiribella}, ``{Perfect discrimination of no-signalling channels via quantum superposition of causal structures},'' {\em Phys. Rev. A}, vol.~86, p.~040301, Oct. 2012.

\bibitem{Araujo2014}
M.~Ara\'ujo, F.~Costa, and {\v{C}}.~Brukner, ``Computational advantage from quantum-controlled ordering of gates,'' {\em Phys. Rev. Lett.}, vol.~113, p.~250402, Dec 2014.

\bibitem{Guerin2016}
P.~A. Gu\'erin, A.~Feix, M.~Ara\'ujo, and {\v C}.~Brukner, ``Exponential communication complexity advantage from quantum superposition of the direction of communication,'' {\em Phys. Rev. Lett.}, vol.~117, p.~100502, Sep 2016.

\bibitem{OreshkovGiarmatzi_2016}
O.~Oreshkov and C.~Giarmatzi, ``Causal and causally separable processes,'' {\em New Journal of Physics}, vol.~18, no.~9, p.~093020, 2016.

\bibitem{costa2016}
F.~Costa and S.~Shrapnel, ``Quantum causal modelling,'' {\em New J. of Phys.}, vol.~18, no.~6, p.~063032, 2016.

\bibitem{Allen2016}
J.-M.~A. Allen, J.~Barrett, D.~C. Horsman, C.~M. Lee, and R.~W. Spekkens, ``Quantum common causes and quantum causal models,'' {\em Phys. Rev. X}, vol.~7, p.~031021, Jul 2017.

\bibitem{QCausal2019}
J.~Barrett, R.~Lorenz, and O.~Oreshkov, ``Quantum causal models,'' 2019.

\bibitem{zych2019}
M.~Zych, F.~Costa, I.~Pikovski, and {\v C}.~Brukner, ``Bell's theorem for temporal order,'' {\em Nature Communications}, vol.~10, p.~3772, 08 2019.

\bibitem{QCausalDiscoveryAlgorithm}
C.~Giarmatzi and F.~Costa, ``A quantum causal discovery algorithm,'' {\em npj Quantum Information}, vol.~4, Mar. 2018.

\bibitem{parker2021background}
L.~Parker and F.~Costa, ``Background independence and quantum causal structure,'' {\em {Quantum}}, vol.~6, p.~865, 2022.

\bibitem{Costa2025}
F.~Costa, J.~Barrett, and S.~Shrapnel, ``A de {F}inetti theorem for quantum causal structures,'' vol.~9, p.~1628.

\bibitem{QComputationsWODefCStruc}
G.~Chiribella, G.~M. D'Ariano, P.~Perinotti, and B.~Valiron, ``Quantum computations without definite causal structure,'' {\em Phys. Rev. A}, vol.~88, p.~022318, Aug 2013.

\bibitem{Araujo_2015}
M.~Ara\'ujo, C.~Branciard, F.~Costa, A.~Feix, C.~Giarmatzi, and {\v C}.~Brukner, ``Witnessing causal nonseparability,'' {\em New Journal of Physics}, vol.~17, p.~102001, oct 2015.

\bibitem{Costa2020}
F.~Costa, ``A no-go theorem for superpositions of causal orders,'' {\em {Quantum}}, vol.~6, p.~663, 2022.

\bibitem{Baumeler2014PRA}
A.~Baumeler, A.~Feix, and S.~Wolf, ``Maximal incompatibility of locally classical behavior and global causal order in multiparty scenarios,'' {\em Phys. Rev. A}, vol.~90, p.~042106, Oct 2014.

\bibitem{Baumeler_2016}
{\"A}.~Baumeler and S.~Wolf, ``The space of logically consistent classical processes without causal order,'' {\em New Journal of Physics}, vol.~18, p.~013036, jan 2016.

\bibitem{Baumeler2022}
{\"{A}}.~Baumeler, A.~S. Gilani, and J.~Rashid, ``Unlimited non-causal correlations and their relation to non-locality,'' {\em {Quantum}}, vol.~6, p.~673, Mar. 2022.

\bibitem{StateDiscrimination}
S.~M. Barnett and S.~Croke, ``Quantum state discrimination,'' {\em Adv. Opt. Photon.}, vol.~1, pp.~238--278, Apr 2009.

\bibitem{UnitaryDiscrim}
R.~Duan, Y.~Feng, and M.~Ying, ``Perfect distinguishability of quantum operations,'' {\em Phys. Rev. Lett.}, vol.~103, p.~210501, Nov 2009.

\bibitem{CombDiscrimination}
K.~Nakahira and K.~Kato, ``Ultimate limits to quantum process discrimination,'' 2020.

\bibitem{Oeckl2003318}
R.~Oeckl, ``A “general boundary” formulation for quantum mechanics and quantum gravity,'' {\em Phys. Lett. B}, vol.~575, no.~3–4, pp.~318--324, 2003.

\bibitem{Chiribella2008supermap}
G.~{Chiribella}, G.~M. {D'Ariano}, and P.~{Perinotti}, ``{Transforming quantum operations: Quantum supermaps},'' {\em EPL (Europhysics Letters)}, vol.~83, p.~30004, Aug. 2008.

\bibitem{Bisio2019}
A.~{Bisio} and P.~{Perinotti}, ``{Theoretical framework for higher-order quantum theory},'' {\em Proceedings of the Royal Society of London Series A}, vol.~475, p.~20180706, May 2019.

\bibitem{Aharonov2009}
Y.~Aharonov, S.~Popescu, J.~Tollaksen, and L.~Vaidman, ``Multiple-time states and multiple-time measurements in quantum mechanics,'' {\em Phys. Rev. A}, vol.~79, p.~052110, 2009.

\bibitem{Silva2017}
R.~Silva, Y.~Guryanova, A.~J. Short, P.~Skrzypczyk, N.~Brunner, and S.~Popescu, ``Connecting processes with indefinite causal order and multi-time quantum states,'' {\em New Journal of Physics}, vol.~19, p.~103022, oct 2017.

\bibitem{Cotler2016}
J.~Cotler and F.~Wilczek, ``Entangled histories,'' {\em Physica Scripta}, vol.~2016, no.~T168, p.~014004, 2016.

\bibitem{Cotler2017}
J.~Cotler, C.-M. Jian, X.-L. Qi, and F.~Wilczek, ``Superdensity operators for spacetime quantum mechanics,'' {\em J. High Energ. Phys.}, vol.~2018, p.~93, sep 2018.

\bibitem{NielsenAndChuang}
M.~Nielsen and I.~Chuang, {\em Quantum Computation and Quantum Information}.
\newblock Cambridge University Press, 10~ed., 2010.

\bibitem{Hardy2012}
L.~Hardy, ``The operator tensor formulation of quantum theory,'' {\em Philos. Trans. Royal Soc. A}, vol.~370, no.~1971, pp.~3385--3417, 2012.

\bibitem{Araujo2017purification}
M.~Ara{\'{u}}jo, A.~Feix, M.~Navascu{\'{e}}s, and {\v{C}}.~Brukner, ``A purification postulate for quantum mechanics with indefinite causal order,'' {\em {Quantum}}, vol.~1, p.~10, Apr. 2017.

\bibitem{Pollock2018}
F.~A. Pollock, C.~Rodr{\' i}guez-Rosario, T.~Frauenheim, M.~Paternostro, and K.~Modi, ``{Operational Markov Condition for Quantum Processes},'' {\em Phys. Rev. Lett.}, vol.~120, p.~040405, 2018.

\bibitem{Pollock2018a}
F.~A. Pollock, C.~Rodr\'{\i}guez-Rosario, T.~Frauenheim, M.~Paternostro, and K.~Modi, ``Non-markovian quantum processes: Complete framework and efficient characterization,'' {\em Phys. Rev. A}, vol.~97, p.~012127, Jan 2018.

\bibitem{lewandowskastrategies2022}
P.~Lewandowska, {\L}.~Pawela, and Z.~Puchała, ``Strategies for single-shot discrimination of process matrices,'' Oct. 2022.

\bibitem{delahamette2022quantum}
A.-C. de~la Hamette, V.~Kabel, M.~Christodoulou, and {\v C}.~Brukner, ``Indefinite causal order and quantum coordinates,'' {\em Phys. Rev. Lett.}, vol.~135, p.~141402, Oct 2025.

\bibitem{Stachel2014}
J.~Stachel, ``The hole argument and some physical and philosophical implications,'' {\em Living Reviews in Relativity}, vol.~17, p.~1, Feb. 2014.

\bibitem{Helstrom1969}
C.~W. Helstrom, ``Quantum detection and estimation theory,'' {\em Journal of Statistical Physics}, vol.~1, no.~2, pp.~231--252, 1969.

\bibitem{WitnessIntro}
C.~Branciard, ``Witnesses of causal nonseparability: an introduction and a few case studies,'' 2016.

\bibitem{Wechs2019}
J.~Wechs, A.~A. Abbott, and C.~Branciard, ``On the definition and characterisation of multipartite causal (non)separability,'' {\em New J. of Phys.}, vol.~21, p.~013027, jan 2019.

\bibitem{MarkovProcesses}
C.~Giarmatzi and F.~Costa, ``Witnessing quantum memory in non-{M}arkovian processes,'' {\em {Quantum}}, vol.~5, p.~440, Apr. 2021.

\bibitem{QHypothesisTesting}
J.~P. Tej, S.~R. Ahmed, A.~R.~U. Devi, and A.~K. Rajagopal, ``Quantum hypothesis testing and state discrimination,'' 2018.

\bibitem{Chiribella2019a}
G.~Chiribella and D.~Ebler, ``Quantum speedup in the identification of cause--effect relations,'' {\em Nature Communications}, vol.~10, no.~1, p.~1472, 2019.

\bibitem{Kundu2022}
A.~Kundu, T.~Acharya, and A.~Sarkar, ``A scalable quantum gate-based implementation for causal hypothesis testing,'' {\em arXiv:2209.02016 [quant-ph]}, 2022.

\end{thebibliography}

\end{document}